\documentclass[useAMS,usenatbib,twocolumn]{mn2e}
\usepackage{amsfonts}
\usepackage{amsmath}
\usepackage[dvips]{graphicx}
\usepackage{epsfig}
\usepackage{wrapfig}
\usepackage{amssymb}
\usepackage{longtable}
\usepackage{accents}
\usepackage{enumitem} 
\SetLabelAlign{CenterWithParen}{\hfil({#1})\hfil}  
\usepackage{dcolumn}
\usepackage{multirow}
\usepackage{bm}
\usepackage[caption=false]{subfig}

\begin{document}

\title[The oscillatory motion of the jet of S5 1803+784]{Flaring radio lanterns along the ridge line: long-term oscillatory motion in the jet of S5~1803+784}
\author[E. Kun, M. Karouzos, K. \'{E}. Gab\'{a}nyi, S. Britzen, O. M. Kurtanidze, L. \'{A}.
Gergely]
{\parbox{\textwidth}{E. Kun$^{1}$\thanks{%
E-mail: kun@titan.physx.u-szeged.hu}, M.
Karouzos$^{2}$, K. \'{E}. Gab\'{a}nyi$^{3,4}$, S. Britzen$^{5}$,  O. M. Kurtanidze$^{6,7,8}$,\\ L. \'{A}. Gergely$^{1}$}\\
\vspace{0.4cm}\\
$^{1}$Institute of Physics, University of Szeged, D\'om t\'er 9, H-6720 Szeged, Hungary\\
$^{2}$Nature Astronomy, Springer Nature, 4 Crinan Street, N1 9XW, London, UK\\
$^{3}$Konkoly Observatory, MTA Research Centre for Astronomy and Earth Sciences, P.O. Box 67, H-1525 Budapest, Hungary\\
$^{4}$MTA-ELTE Extragalactic Astrophysics Research Group, ELTE TTK P\'{a}zm\'{a}ny P\'{e}ter s\'{e}t\'{a}ny 1/A, H-1117, Budapest, Hungary\\
$^{5}$Max-Planck-Institute f\"{u}r Radioastronomie, Auf dem H\"{u}gel 69, D-53121 Bonn, Germany\\
$^{6}$Abastumani Observatory, Mt. Kanobili, Abastumani 0301, Georgia\\
$^{7}$Engelhardt Astronomical Observatory, Kazan Federal University, Tatarstan, Russia\\
$^{8}$Landessternwarte, Zentrum für Astronomie der Universität Heidelberg,Königstuhl 12, 69117 Heidelberg, Germany}

\date{Accepted . Received ; in original form }
\maketitle

\begin{abstract}
We present a detailed analysis of $30$ very long baseline interferometric observations of the BL Lac object S5 1803+784 ($z=0.679$), obtained between mean observational time $1994.67$ and $2012.91$ at observational frequency 15 GHz. The long-term behaviour of the jet ridge line reveals the jet experiences an oscillatory motion superposed on its helical jet kinematics on a time-scale of about 6 years. The excess variance of the positional variability indicates the jet components being farther from the VLBI core have larger amplitude in their position variations. The fractional variability amplitude shows slight changes in $3$-year bins of the component's position. The temporal variability in the Doppler boosting of the ridge line results in jet regions behaving as flaring "radio lanterns". We offer a qualitative scenario leading to the oscillation of the jet ridge line, that utilizes the orbital motion of the jet emitter black hole due to a binary black hole companion. A correlation analysis implies composite origin of the flux variability of the jet components, emerging due to possibly both the evolving jet-structure and its intrinsic variability.
\end{abstract}

\pagerange{\pageref{firstpage}--\pageref{lastpage}}

\label{firstpage}


\section{Introduction}

Accretion onto a highly-spinning supermassive black hole (SMBH) powers relativistic outflows, also known as jets, emanating from the compact centre of the active galactic nuclei \citep[AGN,][]{BlandfordZnajek1977,Begelman1980}. The jets are thought to originate in the vicinity of the SMBH, therefore in principle they are able to carry away information about the central engine of the AGN. Understanding the process of the jet-launching is important to probe the AGN and its central black hole (BH). The improved resolution of the forthcoming Event Horizon observations on the shadow of the central black hole of the Galaxy promises important updates on the supermassive black hole-phenomena \citep[e.g.][]{Gold2016}.

Very long baseline interferometric (VLBI) radio observations of jets resolve their surface brightness distribution into several distinct components. The origin of these components is still unclear, whether they are shocks in the plasma \citep[e.g.,][]{Marscher1985}, or strongly Doppler-boosted regions \citep[e.g.][]{Camenzind1992}. With the advent of large monitoring programs, the structural evolution of many jets is observed on time-scales of years, even decades. A prime example is the Monitoring Of Jets in Active galactic nuclei with VLBA Experiments \citep[MOJAVE,][]{Lister2009,Lister2013} programme. The MOJAVE database provides access to radio data of extragalactic jets since $1994.64$, obtained with the Very Long Baseline Array (VLBA) at the $15$ GHz observing frequency. The statistically-complete sample of the radio-selected active galaxies from the Caltech-Jodrell Bank flat-spectrum (CJF) sample was also extensively explored \citep[e.g.][]{Vermeulen2003,Pollack2003,CJF1Britzen2007,CJF2Britzen2008}. 

In this paper we analyse the calibrated $uv$-data of the jet of blazar S5 1803+784 \citep[$z=0.679$,][]{Lawrence1996}, employing MOJAVE observations. In contrast to what has been observed in the majority of VLBI jets, its components do not exhibit outward motion, but rather are oscillating about certain positions of the jet \citep{Britzen2005b,Britzen2005a,Britzen2010}. S5 1803+784 exhibits a $76\degr$ difference in position angle between the parsec and kiloparsec-scale jet \citep[][and references therein]{Britzen2005a}. The transition region was mapped by \citet{Britzen2005b} using $1.5$, $5$, and $15$ GHz VLBI observations. The authors found jet wiggles on all scales between $1$ mas to $50$ arcsec, suggesting an oscillation on $50$ to $100$ kpc scales. \citet{Jorstad2005} found that the high-resolution $43$ GHz images of S5 1803+784 support the hypothesis of a helical structure of the jet. Polarization measurements of S5 1803+784 conducted at 43 GHz reveal high degrees of polarization \citep{Cawthorne2013}. 

\citet{Britzen2010} (hereafter B2010) analysed more than $90$ epochs of radio observations at $6$ frequencies ($1.6$, $2.3$, $5$, $8$, $15$ and $22$ GHz). The core separation and position angle changes of the inner components was found to correlate with the total flux density light curves at $8$ and $15$ GHz. They found that the core separation, the position angle, and the flux density change in a periodic way with a period of $\sim 4$ yr for the inner jet components at $8$ and $15$ GHz. This value is similar to what has been found by \citet{Kelly2003} in the total flux density light curves. B2010 reported that the width of the jet changes with a period of $\sim8.5$ years, and the components exhibit smooth changes in their position angles. They found the most simple model explaining this behaviour is the twisted rotating helix with the period $\sim8.5$ years. This behaviour is qualitatively similar to what is expected from the model of an orbiting jet base presented in \citet{Kun2014}.

The above results motivated us to model-fit VLBI data provided by the MOJAVE team being not included in B2010, and study the evolution of the jet ridge line over a longer time-span. The paper is organised as follows. In Section \ref{long_term_behaviour} we summarize the long-term behaviour of the jet-components in S5 1803+784. In Section \ref{oscillation} we characterize the time-scale of the oscillation of the ridge line of the jet, and the positional variability of its components.
In Section \ref{section:helical_jet} we characterize the helical shape of the jet, and give a set of parameters describing it. In Section \ref{section:orbiting_bh}, we demonstrate the simulation of the jet-structure using the hypothesis of an orbiting BH as the jet emitter, capable to qualitatively explain the oscillatory ridge line of the jet of S5 1803+784. 
In Section \ref{discussion} we discuss the results, and in Section \ref{section:summary} we summarize our findings.

$\Lambda\mathrm{CDM}$ cosmology is used throughout the paper with Hubble constant $H_0=67.8\pm0.9\mbox{~}\mathrm{km\mbox{~}s^{-1}\mbox{~}Mpc^{-1}}$ and matter density $\Omega_\mathrm{m}=0.308\pm0.012$ \citep{Planck2015}. As a result, S5 1803+784 has a luminosity distance of $3684.2$ Mpc and an angular scale of $6.917$~kpc~arcsec$^{-1}$. 

\section{Long-term behaviour of the jet components of S5 1803+784}
\label{long_term_behaviour}

\subsection{Model-fitting of the VLBI data and component identification}

Archival calibrated data of S5 1803+784 at 15 GHz, provided by the MOJAVE programme, cover $18.24$ years. We use the model-fit results of B2010 concerning the first $13$ epochs ($1994.67$--$2005.68$), and supplement it with the next $17$ epochs ($2005.85$--$2012.91$), model-fitted in the present work. Standard \textsc{Difmap} tasks \citep{Shepherd1994} were used to perform the model-fitting on the VLBI data, employing Gaussian components to build up the surface brightness distribution of the jet. To decrease the degrees of freedom during the model-fit and to remain consistent through the epochs, we used only circular Gaussian components to fit the brightness profile of the jet. The summary of the $15$ GHz image parameters is given in Table \ref{table_impars}.

\begin{table*}
\begin{center}
\caption{Summary of the 15 GHz image parameters. (1) epoch of the VLBA
observation, (2) VLBA experiment code, (3)--(4) FWHM minor and major axis of the restoring
beam, respectively, (5) position angle of the
major axis of the restoring beam measured from North to East, (6) rms noise of the image after the model-fit procedure, (7) reduced $\chi{^2}$ of the \textsc{Difmap} model-fit, (8) number of the components in the model.}
\centering
\begin{tabular}{cccccccc}
\hline
\hline
	&	VLBA  & $\mathrm{B}_{\mathrm{min}}$ & $\mathrm{B}_{\mathrm{maj}}$& $\mathrm{B}_{\mathrm{PA}}$ & rms  & Red. & Comp.\\
Epoch & Code & (mas) & (mas) & ($^{\circ}$) & (mJy bm$^{-1}$) & $\chi{^2}$& Number \\
(1) & (2) &  (3) &  (4) &  (5) &  (6) & (7) & (8)\\
\hline
2005-11-07 & BL123P & $0.641$ & $0.648$ & $-6.35$ & $0.24$ & $1.01$ & $7$\\
2006-09-06 & BL137I & $0.617$ & $0.646$ & $82.10$ & $0.42$ & $1.20$ & $7$\\
2007-02-05 & BL137N & $0.479$ & $0.510$ & $-7.77$ & $0.29$ & $1.14$ & $7$\\
2007-05-03 & BK134D & $0.503$ & $0.694$ & $-19.94$ & $0.17$ & $1.17$ & $9$\\
2007-08-16 & BL149AF & $0.618$ & $0.681$ & $-31.13$ & $0.21$ & $1.35$ & $7$\\
2008-08-25 & BL149BB & $0.579$ & $0.612$ & $-16.69$ & $0.22$ & $1.35$ & $7$\\
2008-12-05 & BG192 & $0.503$ & $0.727$ & $11.09$ & $0.22$ & $2.09$ & $7$\\
2009-03-25 & BL149BJ & $0.635$ & $0.720$ & $16.14$ & $0.17$ & $1.20$ & $7$\\
2009-10-27 & BL149CC & $0.631$ & $0.673$ & $-1.92$ & $0.16$ & $1.13$ & $7$\\
2010-02-11 & BL149CH & $0.627$ & $0.701$ & $43.02$ & $0.20$ & $1.10$ & $10$\\
2010-12-18 & BM300 & $0.568$ & $0.752$ & $22.24$ & $0.17$ & $0.81$ & $10$\\
2011-03-05 & BL149DD & $0.756$ & $0.833$ & $53.78$ & $0.17$ & $1.22$ & $8$\\
2011-05-26 & BL149DI & $0.600$ & $0.672$ & $-28.86$ & $0.18$ & $1.13$ & $8$\\
2011-12-29 & BL178AD & $0.656$ & $0.726$ & $20.23$ & $0.20$ & $1.46$ & $6$\\
2012-03-04 & BL178AH & $0.685$ & $0.711$ & $44.68$ & $0.16$ & $1.44$ & $7$\\
2012-05-24 & BL178AK & $0.616$ & $0.691$ & $-4.51$ & $0.20$ & $1.36$ & $6$\\
2012-11-28 & BL178AU & $0.487$ & $0.516$ & $-15.74$ & $0.54$ & $1.21$ & $7$\\
\hline
\hline
\label{table_impars}
\end{tabular}
\end{center}
\end{table*}

\begin{table*}
\begin{center}
\caption{Circular Gaussian model-fit results for S5 1803+784. (1) epoch of observation, (2) flux density, (3)-(4) position of the component center respect to the core, (5) FWHM major axis, (6) jet-component identification. The full table is available in electronic format online as Supporting Information.}
\begin{tabular}{lccccc}
\hline
\hline
Epoch & Flux density & r & $\theta$ & d  & CO\\ 
(yr) & (Jy) & (mas) & (deg) & (mas)  & \\ 
(1) & (2) & (3) & (4) & (5)  & (6)\\ 
\hline
$	2005.85	$	&	$	1.33	\pm	0.14	$	&	$	0.00	\pm	0.00	$	&	$	0.00	\pm	0.00	$	&	$	0.08	\pm	0.01	$	&	Cr\\
	&	$	0.17	\pm	0.02	$	&	$	0.25	\pm	0.02	$	&	$	-75.03	\pm	3.00	$	&	$	0.23	\pm	0.04	$	&	C0\\			
	&	$	0.05	\pm	0.01	$	&	$	0.78	\pm	0.02	$	&	$	-76.93	\pm	0.98	$	&	$	0.27	\pm	0.03	$	&	C1\\			
	&	$	0.22	\pm	0.02	$	&	$	1.43	\pm	0.02	$	&	$	-90.72	\pm	0.03	$	&	$	0.26	\pm	0.01	$	&	Ca\\			
	&	$	0.09	\pm	0.01	$	&	$	1.75	\pm	0.03	$	&	$	-88.40	\pm	0.09	$	&	$	0.56	\pm	0.03	$	&	C2\\			
	&	$	0.03	\pm	0.00	$	&	$	3.81	\pm	0.10	$	&	$	-85.17	\pm	0.59	$	&	$	1.97	\pm	0.18	$	&	C4\\			
	&	$	0.06	\pm	0.00	$	&	$	6.50	\pm	0.14	$	&	$	-96.98	\pm	2.25	$	&	$	2.69	\pm	0.10	$	&	C8\\	
\hline
\hline
\label{longtable}
\end{tabular}
\end{center}
\end{table*}

We cross-identified components across different epochs by requiring a smooth change for core separation, flux density, and full width at half maximum (FWHM) of the fitted Gaussian. We identify nine components: Cr, C$0$, C$1$, C$a$, C$2$, B$3$, C$4$, C$8$, C$12$, where the labelling follows the notations of B2010. Increasing numbers indicate larger distances to the VLBI core, Ca being situated between components C1 and C2. The easternmost and brightest component, denoted by Cr, is assumed to be the VLBI core, such as in the earlier works carried out at the 15 GHz observing frequency \citep[][]{Roland2008,Britzen2010}. The best-fit values of the integrated flux density, core separation, position angle and full width at half maximum (FWHM) size of the fitted jet components are given in Table \ref{longtable}. Error-estimation of these parameters was performed in the same way as in \citet{Kun2014}.

\subsection{Core separations, proper motions}

\begin{figure*}
\centering
    \includegraphics[width=0.8\textwidth]{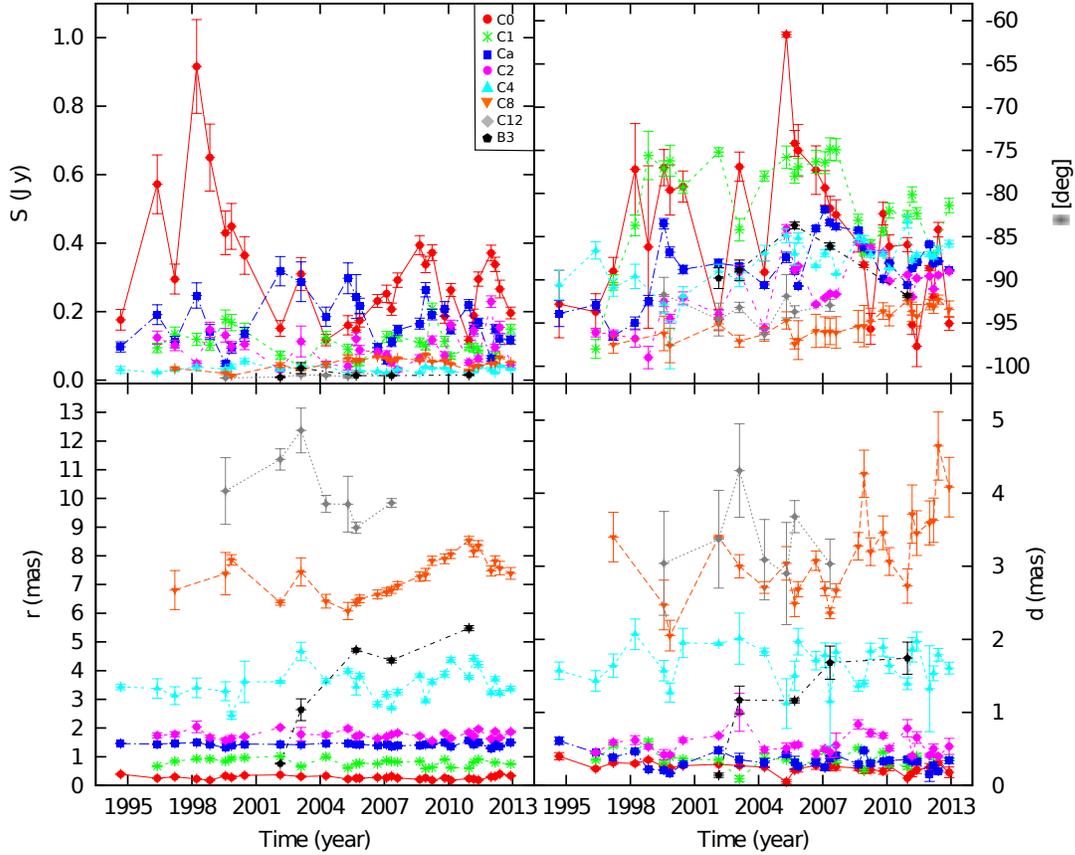}
    \caption{Flux density (S), core separation (r), position angle ($\theta$) and width ($d$) of C0, C1, Ca, C2, C4, C8, C12 plotted against time.}
    \label{fig:coresep_all}
\end{figure*}

The integrated flux density, position angle, core separation and FWHM-width of the VLBI jet components as a function of time are presented in Fig. \ref{fig:coresep_all}. The third panel of this figure clearly shows the jet components maintain quasi-statinoary core separations during the observations, rather then exhibiting global outward motion. The average values of the core separations are listed in Table \ref{table:velocities}. The model-fit results of B2010 proved the presence of one fast moving component (B3) in addition to the stationary components. B3 has been observed between 2002.1 and 2005.7 and shows a kinematical behaviour that differs significantly from the other components. We identified component B$3$ at two additional epochs ($2007.34$, $2010.96$) based on our component-identification strategy. The fact that this component exhibits outward superluminal motion suggests that the launching scenario and the propagation of B$3$ might be driven differently than the rest of the jet features. We will discuss its possible nature in Section \ref{discussion}.

Compared to the measured proper motions of the jet components in B2010, all components have smaller proper motion velocities when considering the total sample. In particular, the apparent speed of the inner jet components (C0, C1, Ca, C2) do not exceed $0.01$ mas yr$^{-1}$. The velocity of C4 drops to approximately 10 per-cent of the velocity from B2010, with significance $>7\sigma$. The proper motion of component C8 of the outer jet changes its sign from positive to negative when all 30 epochs are considered. If the jet is indeed experiencing an oscillatory motion, it is expected that the jet component speed converges to zero when measured over longer periods of time.

\subsection{Evolution of the jet width}
\label{section:evoljetridgeline}

\begin{figure*}
    \includegraphics[scale=0.8]{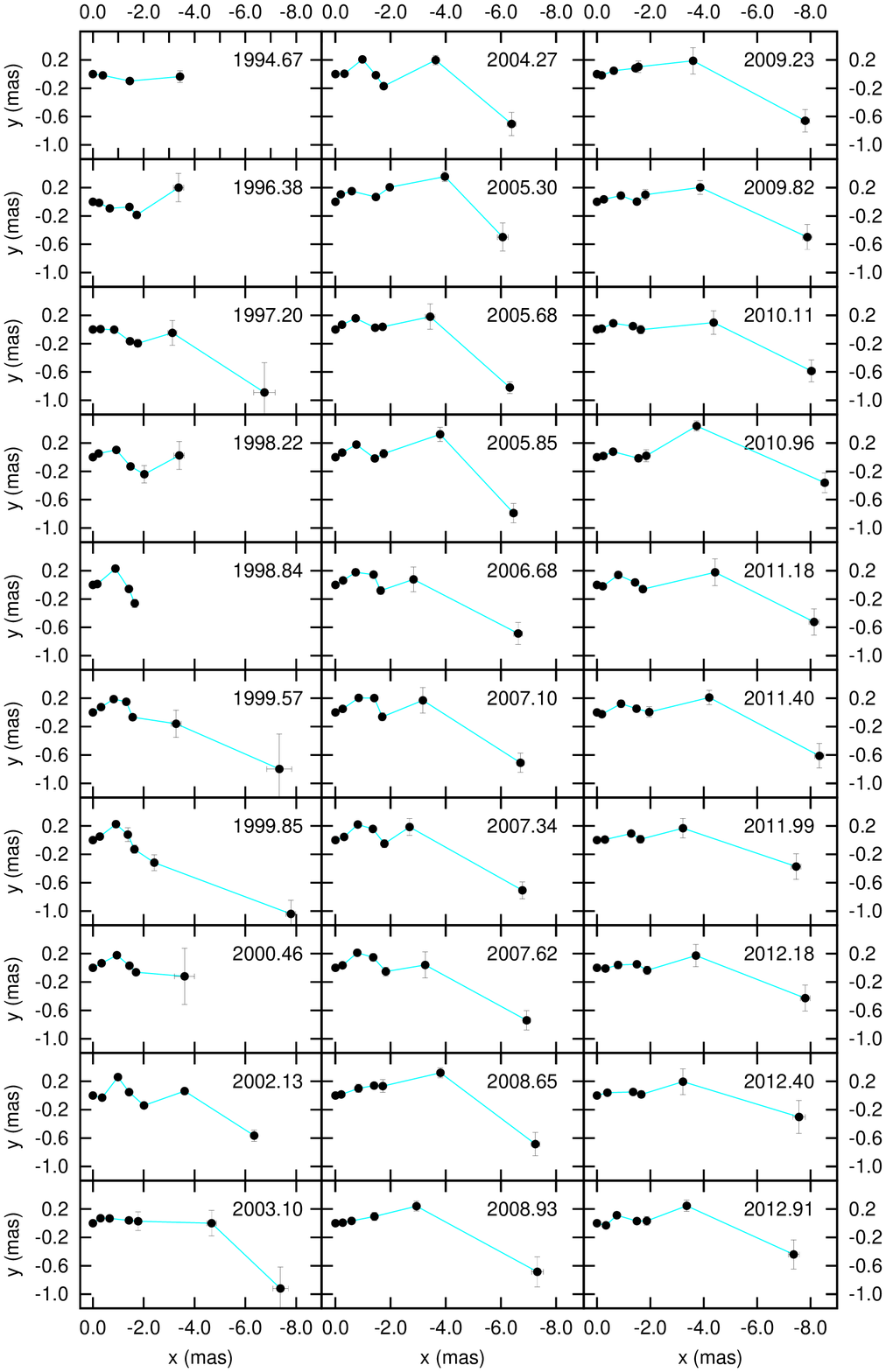}
    \caption{$xy$-coordinate maps of the jet of S5 1803+784 at 15 GHz, taking the core and components C$0$, C$1$, C$a$, C$2$, C$4$, C$8$ into account. Observing epochs can be found in the upper right corner of the maps.}
    \label{fig:allxy}
\end{figure*}

\begin{figure}
    \includegraphics[scale=0.7]{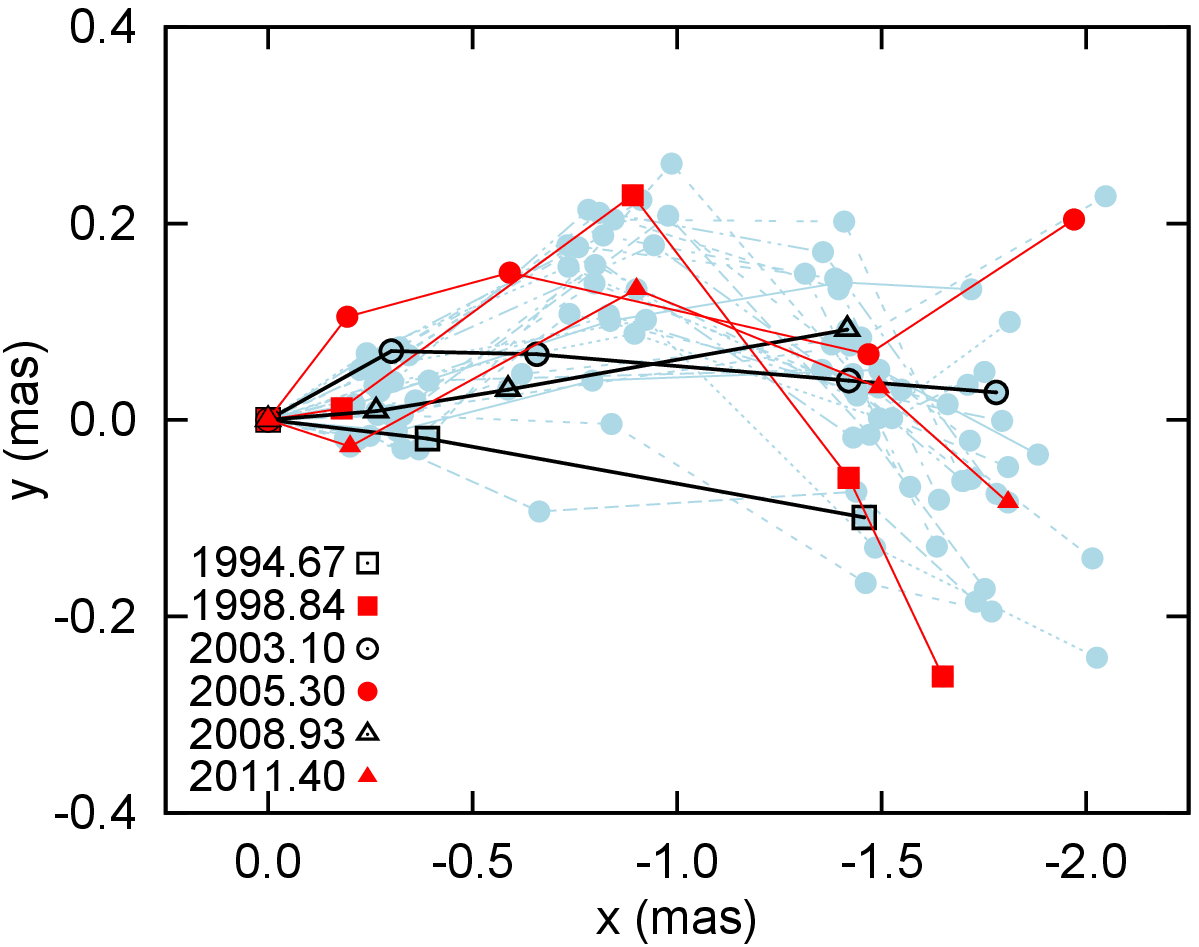}
    \caption{$xy$-position variability of the components of the inner jet. We show three examples of the straight (black lines and symbols) and of the sinusoidally bent ridge lines (red line and symbols), superposed on $xy$-positions from all observing epochs (light blue lines and filled circles.}
    \label{fig:focus}
\end{figure}

The intrinsic jet motion can be studied by defining the jet ridge line, a line that connects the centres of all VLBI jet components identified at a single epoch of observations \citep[e.g.][]{Hummel1992,Britzen2010,Karouzos2012a}. We plot the jet ridge line across all available epochs at $15$ GHz in Fig. \ref{fig:allxy}. We only consider the core and those identified jet-features, that appear at least in $10$ epochs (C$0$, C$1$, C$a$, C$2$, C$4$, C$8$). The ridge line of the inner jet is shown in Fig. \ref{fig:focus} (C$0$, C$1$, C$a$, C$2$), with three examples of the straight, and three examples of the curved jet-structure. It seems the ridge line of the inner jet evolves between straight and curved shapes, underlining the finding of B2010. 

To characterize the transitions between the sinusoidally bent and straight shapes of the jet, we calculated the apparent width of the jet \citep{Karouzos2012a,Karouzos2012b}, characterizing the observed opening angle of the jet, defined as \citep{Karouzos2012b}:
\begin{equation}
dP_\mathrm{i}=\theta_\mathrm{i}^\mathrm{max}-\theta_\mathrm{i}^\mathrm{min},
\end{equation}
where $\theta_\mathrm{i}^\mathrm{max}$ and $\theta_\mathrm{i}^\mathrm{min}$ are the maximal and minimal position angles measured in the $i$th epoch, respectively. We calculated the apparent width by excluding C$4$ and C$8$, in order to focus on the inner jet, where the most violent changes are found.

\begin{figure}
\includegraphics[scale=0.45]{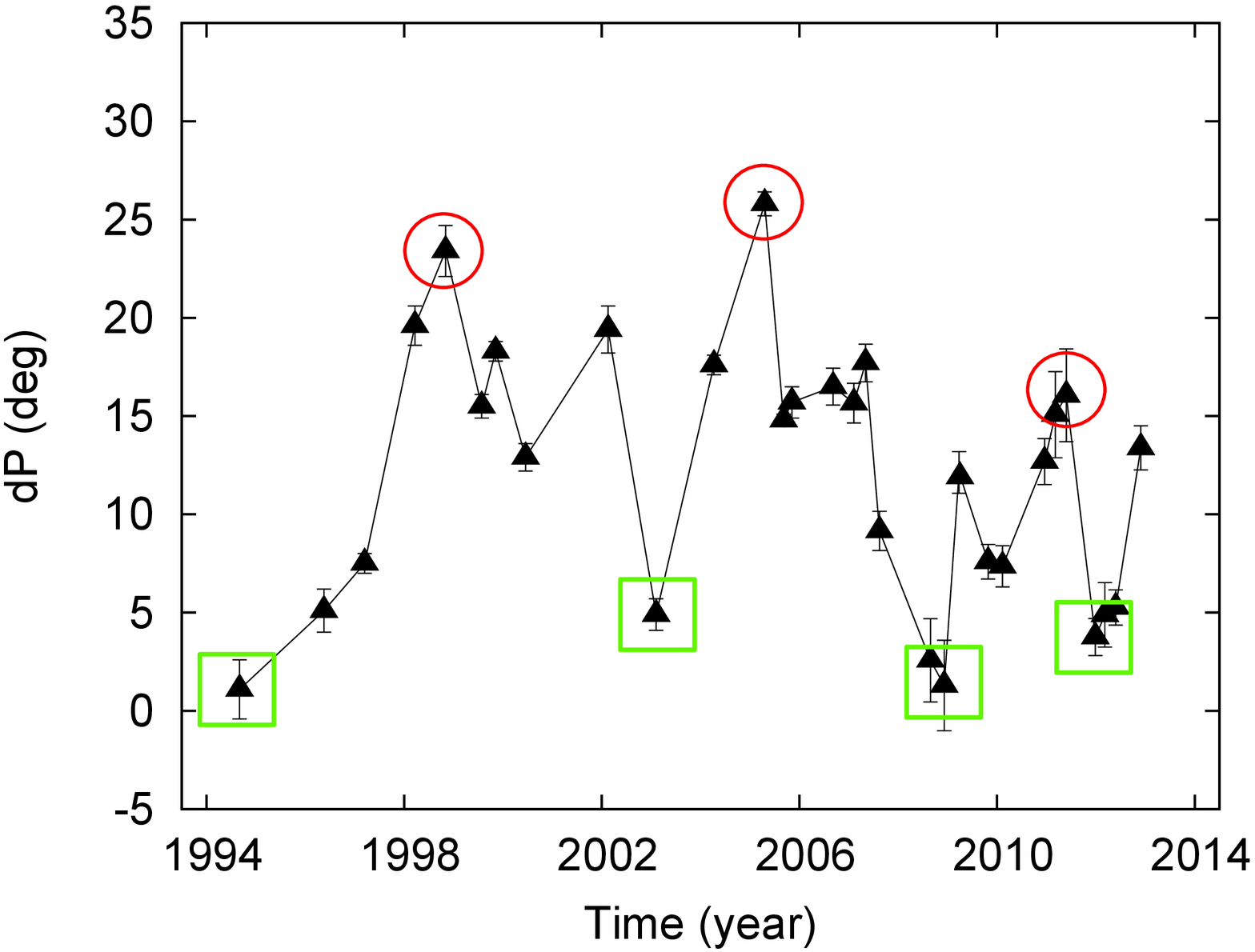}
\caption{Apparent width of the jet taking components C$0$, C$1$, C$a$, C$2$ into account. The epochs when the jet was in curved (straight) mode are marked by red circles (green boxes).}
\label{fig:appwidth}
\end{figure}

In Fig. \ref{fig:appwidth} we show the time variation of the jet-width, marking the straight and curved modes. The jet was straight in $t_\mathrm{s,1}=1994.67$, $t_\mathrm{s,2}=2003.10$, $t_\mathrm{s,3}=2008.93$, $t_\mathrm{s,4}=2011.99$, and curved in $t_\mathrm{c,1}=1998.84$, $t_\mathrm{c,2}=2005.30$, $t_\mathrm{c,3}=2011.40$. Averaging the differences between the subsequent epochs of one mode (straight versus curved), such that:
\begin{flalign}
P_\mathrm{w}=\frac{1}{N-2} \left[ \sum \limits_\mathrm{i=1}^{3} (t_\mathrm{s,i+1}-t_\mathrm{s,i}) + \sum\limits_\mathrm{i=1}^{2} (t_\mathrm{c,i+1}-t_\mathrm{c,i}) \right],
\end{flalign}
where $N=7$ is the number of epochs in question. Substituting epochs $t_{s}$ and $t_{c}$, we conclude that the inner jet changes its width with a period of $P_\mathrm{w}\sim6$ yr. We calculate the apparent half-opening angle of the jet as half the average width of the inner jet, emerging as $\psi^\mathrm{obs}=6.05\degr\pm1.1\degr$.

\begin{table*}
\caption{Linear proper motions and apparent velocities. (1) jet-component identification, (2) average core separation between $1994.67$ and $2012.91$, (3) proper motion between $1994.67$ and $2005.68$ (B2010), (4) proper motion between $2005.85$ and $2012.91$, (5) apparent speed between $1994.67$ and $2012.91$.}
\label{table:velocities}
\begin{tabular}{lcccc}
\hline
\hline
Comp. ID & $r_\mathrm{av}$ & $\mu_r^{B2010}$& $\mu_r^\mathrm{15GHz,all}$ & $\beta_\mathrm{app,r}$\\
 & (mas) & (mas yr$^{-1}$) & (mas yr$^{-1}$) &  ($c$)\\
(1) & (2) & (3) & (4) &  (5)\\
\hline
C0 & $0.28 \pm 0.02$ & $0.017 \pm 0.001$ & $-0.000 \pm 0.002$ & $-0.02 \pm 0.08$\\
C1 & $0.80 \pm 0.03$ & $- 0.037 \pm 0.003$ & $-0.009 \pm 0.006$ & $-0.34 \pm 0.23$\\
Ca & $1.43 \pm 0.02$ & $0.007 \pm 0.004$ & $-0.001 \pm 0.002$ & $-0.05 \pm 0.08$\\
C2 & $1.76 \pm 0.06$ & $0.017 \pm 0.007$ & $-0.008 \pm 0.006$ & $-0.32 \pm 0.23$\\
C4 & $3.53 \pm 0.16$ & $0.145 \pm 0.017$ & $0.018 \pm 0.022$ & $0.68 \pm 0.82$\\
C8 & $7.29 \pm 0.23$ & $- 0.182 \pm 0.005$ & $0.141 \pm 0.033$ & $5.33 \pm 1.23$\\
C12 & $10.35 \pm 0.56$ & $- 0.715 \pm 0.113$ & $-0.186 \pm 0.165$ & $-7.03 \pm 6.25$\\
B3 & $3.59 \pm 0.13$ & $0.807 \pm 0.151$ & $0.669 \pm 0.139$ & $25.34 \pm 5.26$\\
\hline
\hline
\end{tabular}
\end{table*}

\section{Oscillation}
\label{oscillation}
\subsection{Characteristic time-scale of the oscillation}
\label{minima}

\begin{figure}
   \includegraphics[scale=0.65]{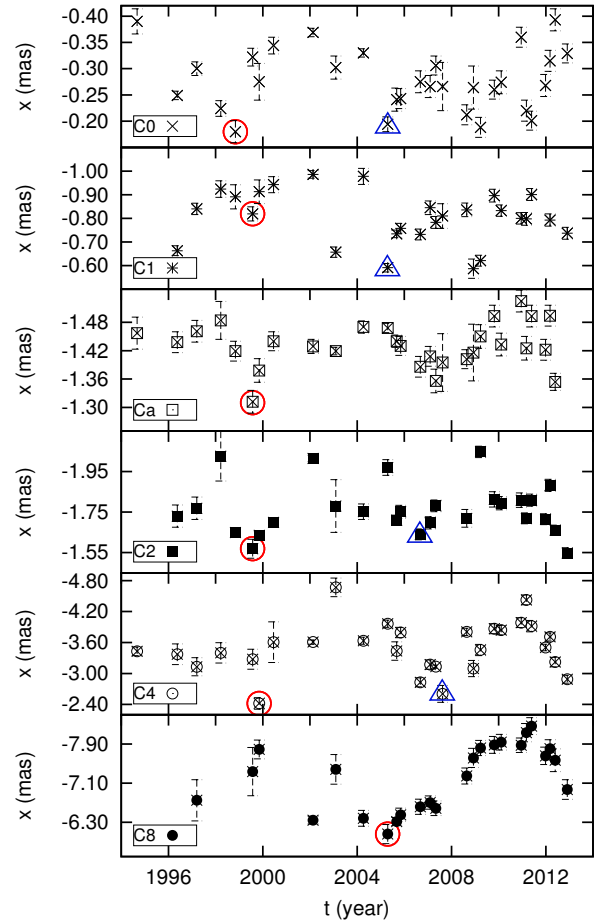}
    \caption{Time variability of the $x$-coordinate of components detected in at least $10$ epochs (C$0$, C$1$, C$a$, C$2$, C$4$, C$8$). Red empty circles denote $x$-minima from the $1$st sequence, and blue empty triangles denote $x$-minima from the $2$nd sequence.}
    \label{fig:r_all}
\end{figure}

The jet of S5 1803+784 has an almost perfectly east--west orientation, therefore the variations of the relative declination (Dec, $y$-coordinate relative to the VLBI core) of the components are minor compared to those of the relative right ascension (RA, $x$-coordinate relative to the VLBI core). Approximating the separation with the $x$-coordinate only, one can cut the component-errors down to the minimum. The $x$-coordinates of the jet-features are represented in Fig. \ref{fig:r_all} against their observing epochs.


Below we define the local minima of $x$-positions, and focus on their respective epoch. The $m$th feature has its minimum $|x|$-position at the $i$th epoch, if the $|x|$-position at $i$th epoch is smaller than the $|x|$-positions measured at the nearest four epochs:
\begin{flalign}
\min x_{m,i} : (|x| \pm \Delta x)_{m,i\pm 2} &>(|x| \pm \Delta x)_{m,i}\nonumber \\ \&  (|x| \pm \Delta x)_{m,i\pm1} &>(|x| \pm \Delta x)_{m,i},
\end{flalign}
where $\Delta x$ is the $x$-positional error. Problems might arise with this definition if the component data are very sparse in time, but for S5~1803+784 the sampling is fairly uniform. We defined the minima with the four nearest neighbouring coordinates in the time axis, in order to dampen the effect of this kind of bias. Due to this definition, the first two and the last two epochs cannot be considered when finding the minima. It has to be also noted, the model built to estimate the errors on the parameters of the VLBI jet-component only reflects the statistical image errors and the error-estimates are assumed uncorrelated, therefore they should be viewed with caution \citep[see references in][]{Kun2014}.

Using the above we identified $10$ local minima of $|x|$-positions along the jet, marked in Fig. \ref{fig:r_all}. 
We estimated the average time-shift $T_\mathrm{obs}$ between the two minima-sequences. For this we took into account those features, for which there are minima available from both sequences (C$0$, C$1$, C$2$, C$4$). Then if $t_{1,m}$ and $t_{2,m}$ is the epoch of the minimum $|x|$-position of the $m$th feature from the first and second sequences respectively, then $T_\mathrm{obs}=\sum_m^N ({t_{2,m}}-t_{1,m})/N\approx 6.8$ yr ($N=4$). The error on $T_\mathrm{obs}$ was estimated based on the average sampling of the components at those epochs which are used to derive $T_\mathrm{obs}$. Then the time-shift emerged as $T_\mathrm{obs}=6.8\pm0.6$ yr. In Section \ref{section:evoljetridgeline} we concluded that the inner jet varies its width with a period of $\sim6.7$ yr. This value is consistent with the period $T_\mathrm{obs}$ determined from the minima, suggesting they have the same underlying cause.

\subsection{Positional variability}

\begin{figure}
    \includegraphics[scale=0.45]{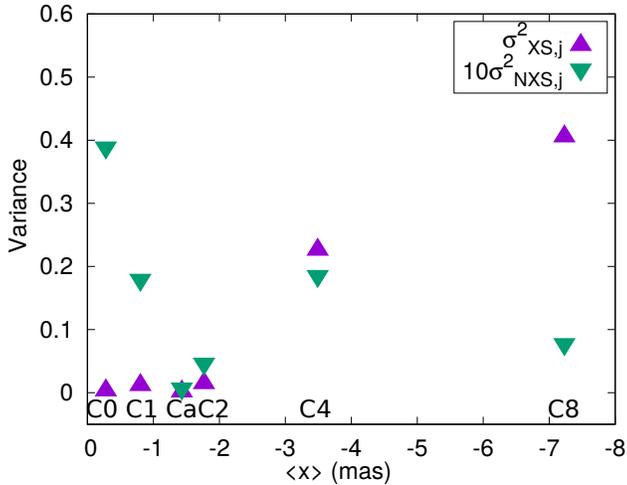}
    \caption{Excess variance $\sigma^2_\mathrm{XS,j}$ (purple up-pointing triangles), normalized excess variance $\sigma^2_\mathrm{NXS,j}$ (green down-pointing  triangles) for 6 jet components as function of the mean $x$-positions $<x>$.}
    \label{fig:variance}
\end{figure}

A stationary process is one for which the statistical properties (such as mean, variance etc.) do not depend on the time. In statistics the variance measures how much a stochastic variable spreads out from its average value; the large variance suggesting significant changes. The excess variance $\sigma^2_\mathrm{XS,j}$, which is the variance after subtracting the expected measurements errors, is a commonly employed quantity to characterize the variability properties of AGN \citep[e.g. of their X-ray light curves][]{Vaughan2003}. We characterize the \textit{intrinsic} variability of the jet components by using the excess variance of their position-dependence on time. For this purpose we consider $x_{i,j}$-coordinates of $j$ components, C0, C1, Ca, C2, C4, C8, measured at the $i$th epoch. Then the excess variance $\sigma^2_\mathrm{XS,j}$ of the $j$th jet component is \citep{Nandra1997,Edelson2002}:
\begin{flalign}
\sigma^2_\mathrm{XS,j}=S_j^2-\bar{\sigma}_\mathrm{err,j}=\frac{1}{N_j-1} \sum\limits_\mathrm{i=1}^{N_j} (x_{i,j}-\bar{x}_j)^2-\frac{1}{N_{j}}\sum\limits_\mathrm{i=1}^{N_j} \sigma_\mathrm{err,j}^2
\end{flalign}
where $S_{j}^2$ is the variance, $\bar{\sigma}_\mathrm{err,j}$ is the mean square of the error of the $x_{i,j}$-positions of the $j$th jet component with mean value $\bar{x}_j$, and $N_j$ is the total number of epochs for the respective component. We define the normalised excess variance as
\begin{equation}
\sigma^2_\mathrm{NXS,j}=\frac{\sigma^2_\mathrm{XS,j}}{\bar{x}^2_j}.
\end{equation}
The results for $\sigma^2_\mathrm{XS,j}$,$\sigma^2_\mathrm{NXS,j}$ are plotted in Fig. \ref{fig:variance}. The excess variance clearly shows the positional variability is function of the average position of the respective jet component, the components being farther from the core showing larger positional variability. It suggests a geometrical effect drives the evolution of the positional variability of the jet-components.

The fractional root mean square variability amplitude $F_\mathrm{var}$ \citep{Edelson1990,Rodriguez1997} can be used to measure whether a measured quantity (e.g. brightness/15-GHz radio flux density of the source/components etc.) show time variations. We use this quantity to characterize the positional variability of the jet components of S5 1803+784 as function of time. We split the $x_j(t)$ curves in three-year bins ($n=6$ bins in total), and calculate $F_\mathrm{var}$ as:
\begin{equation}
F_\mathrm{var}=\sqrt{\frac{S^2-\sigma_{err,j,n}^2}{\bar{x}_{j,n}}},
\end{equation}
where $S^2$ is the total variance of the $x_j(t)$ curve, $\sigma_{err,j,n}^2$ is the mean square of the error of the $x_{i,j}$-positions of the $j$th jet component in the $n$th bin with mean value $\bar{x}_{j,n}$.
Its error is calculated as:
\begin{equation}
F_\mathrm{var,err}=\sqrt{\left[ \sqrt{\frac{1}{2N}}\frac{\sigma_\mathrm{err,j,n}^2}{\bar{x}_{j,n}^2 F_\mathrm{var}}\right]^2+\left[ \sqrt{\frac{\sigma_\mathrm{err,j,n}^2}{N}}\frac{1}{\bar{x}_{j,n}}\right]^2}
\end{equation}

\begin{figure*}
    \includegraphics[scale=0.6]{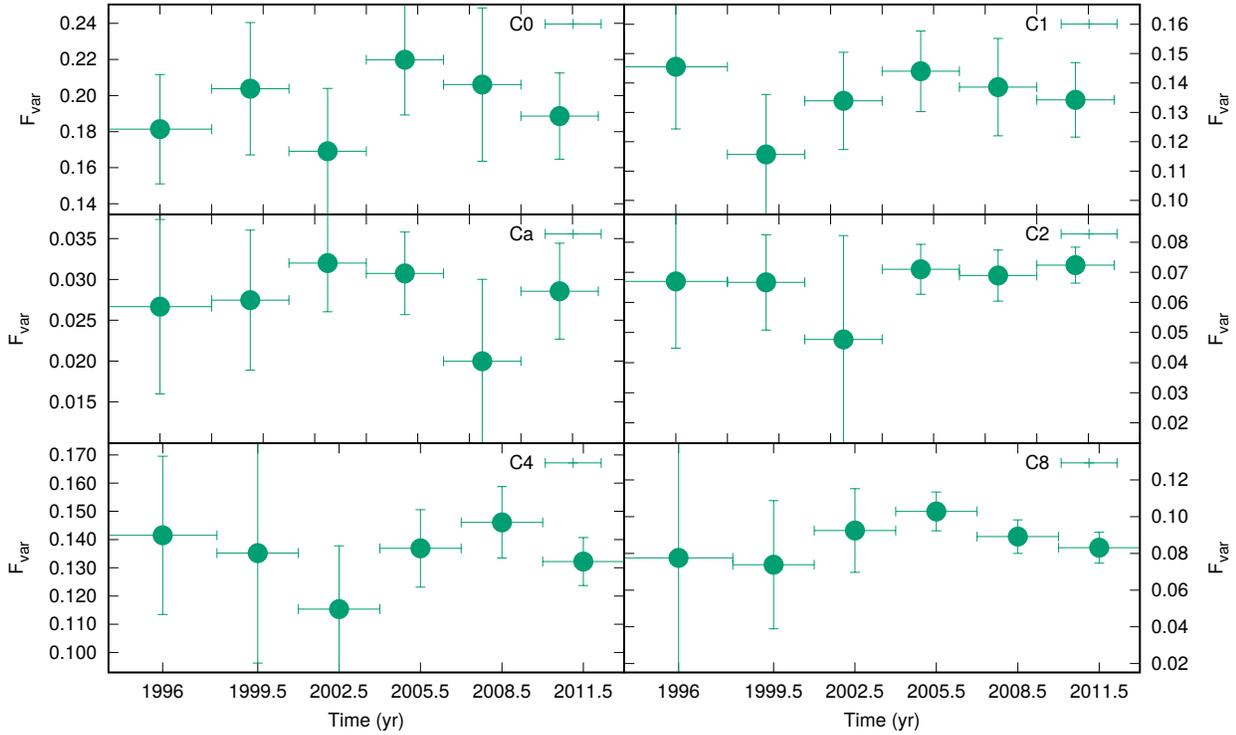}
    \caption{Fractional variability amplitude $F_\mathrm{var}$ for jet components C0, C1, Ca, C2, C4, C8 as function of time in 3-year bins.}
    \label{fig:fvar}
\end{figure*}

The results for $F_\mathrm{var}$ are plotted in Fig. \ref{fig:fvar}. The fractional variability amplitude shows slight changes in $3$-year bins of the component's position, indicating the variable nature of the source.

\section{Helical jet of S5~1803+784}
\label{section:helical_jet}

\subsection{Characteristic spatial orientation of the jet}

\citet{Hovatta2009} calculated the variability Doppler boosting factors, Lorentz factor, and inclination angle for S5~1803+784 to be $\delta\approx12.2$, $\gamma\approx9.5$, $\iota_0\approx4.5$, respectively. A Lorentz factor $\gamma\approx9.5$ indicates jet particles moving outward from the core close to the speed of the light ($\beta_j\approx0.994c$). These values were derived using radio interferometry and optical polarization measurements, occurred before the MOJAVE observations started. As the source is variable, we did not intend to use these values to fix the jet speed and the orientation of the jet flow, instead we calculate these from the MOJAVE measurements employed in this paper, as follows.

For VLBI components, the brightness temperature is calculated as \citep[e.g.][]{Condon1982}:
\begin{flalign}
T_\mathrm{b}=1.22\times 10^{12} (1+z) \frac{S}{d^2 f^2}\mbox{~}\mathrm{K},
\end{flalign}
where $S$~(Jy) is the flux density, $d$~(mas) is the FWHM diameter of the circular Gaussian component, and $f$~(GHz) is the observing frequency. We calculated the \textit{average} brightness temperature of the core as $\bar{T}_\mathrm{VLBI}\approx 1.84\times 10^{12}$~K with standard deviation $1.60\times10^{12}$~K. The large relative value of the latter reflects the variable flux density of the core. Assuming that the intrinsic temperature is equal to the equipartition temperature $5\times 10^{10}$~K \citep[][]{Readhead1994}, the \textit{average} Doppler factor of the core is $\delta_0= 36\pm29$. Assuming that the jet velocity is constant and that B3 reflects its value, then combining $\delta_0$ with the apparent superluminal speed of B3 ($\beta_\mathrm{app}\approx 25.4c$) gives a \textit{characteristic} jet speed as $\beta \approx0.999c$ and \textit{characteristic} inclination as $\iota_0\approx1.7^{\circ}$.

\subsection{Characterization of the jet-shape}

Now we characterize the shape of the jet of S5~1803+784. The previous works motivated the fitting of a helical shape to the component positions \citep{Britzen2001,Tateyama2002,Britzen2005a,Britzen2010,Karouzos2012a}. Conservation laws for the kinetic energy, momentum and jet opening angle lead to a set of equations, that describe how the components move along the jet in cylindrical coordinates \citep[Case 2 of][]{Steffen1995}:
\begin{equation}\label{eq:rt}
r(t)=v_zt \tan \psi + r_0,
\end{equation}
\begin{equation}\label{eq:phi}
\phi(t)=\phi_0 + \frac{\omega_0 r_0}{v_z \tan \psi} \ln \frac{r(t)}{r_0},
\end{equation}
\begin{equation}\label{eq:zt}
z(t)=v_z t,
\end{equation}
where $r$ is the radial distance from the core, $r_0$ is its initial value, $\phi$ is the polar angle, $\phi_0$ is its initial value, $v_z$ is the jet velocity along the $z$-axis which is parallel to the jet symmetry axis, $\psi$ is the half-opening angle of the helix, $\omega_0$ is the initial angular velocity.
The kinematics described by this model is equivalent with the hydrodynamic isothermal helical model of the jet, when the helical mode is not dampened \citep{Hardee1987}. The model results in an increasing spatial wavelength $\lambda (z):=z(\phi+2\pi)-z(\phi)$ (measured along the symmetry axis of the jet) and period $P (t):=t(\phi+2\pi)-t(\phi)$ of the helix, such that:
\begin{flalign}
\lambda(z)&=\left(z+\frac{r_0}{\tan \psi} \right) \left(e^{2\phi \tau} -1\right),\\
P(t)&=\left(t+\frac{r_0}{v_z \tan \psi} \right) \left(e^{2\phi \tau} -1\right),
\end{flalign}
where $\tau=v_z \tan \psi \omega_0^{-1} r_0^{-1}$ is a constant, and depends just on the initial parameters.

In the following, we deduce the helical shape of the jet of S5~1803+784 based on the component positions. The parametrization of the intrinsic jet shape is done as described by Eqs. (\ref{eq:rt}--\ref{eq:zt}). \citet{Steffen1995_1803} noted that there is a small overall curvature in the jet-shape of S5~1803+784. We determined this curvature by fitting a function $f(x)=A x^{B}$, yielding $A=-0.0003 \pm 0.0008$ mas, $B=3.48\pm1.17$, where $x$ is the relative right ascension of the components. We subtracted the best-fit curvature from the original positions to remove this large-scale curvature. Its presence may indicate a reorientation in the jet that happens on far longer time-scales than the VLBI observations, and is therefore not implemented in the model presented here.

\begin{figure*}
    \includegraphics[scale=0.65]{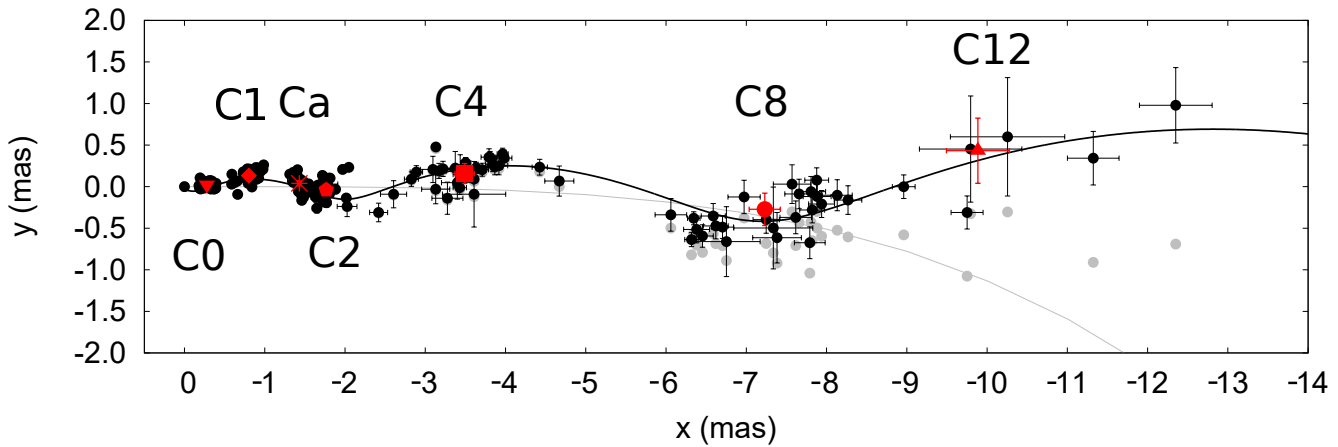}
    \caption{Helical jet-shape of the VLBI jet of S5 1803+784 at the $15$ GHz observing frequency. The original $xy$ component-positions are marked by gray points, along with the fitted and subtracted curvature, marked by the solid gray line. The corrected $xy$-positions of the components are marked by black filled circles with error-bars, while the average $xy$-position of a components are marked by red filled circles. The best-fit helical model is drawn with a solid black line.}
    \label{fig:av_jetshape}
\end{figure*}

The best-fitting parameters of the helical jet model are obtained by iteratively performing non-linear least squares estimates using the Levenberg--Marquardt algorithm, such that the $\chi^2$ was minimized during the process. The fitted parameters were $\omega_0$, $r_0$, $\phi_0$, $\psi$, the fixed parameters were $\beta_{jet}=0.999c$, $\iota_0=1.7^\circ$. We obtain $\omega_0=0.11 \pm 0.02$ mas yr$^{-1}$, $r_0=0.05\pm0.01$ mas, $\phi_0=57.3 \pm 31.0$, $\psi\approx0.05^\circ$. We plot the best-fit model of the helical shape by a single black line in Fig. \ref{fig:av_jetshape}, where we also mark the position of the components, from which the large-scale curvature is already subtracted. 

\section{Orbiting black hole at the jet base}
\label{section:orbiting_bh}

In this section we present the kinematical simulation of a helical jet described by Eqs. (\ref{eq:rt}--\ref{eq:zt}), and show how it responds to the orbiting nature of its base. For this purpose we consider two BHs orbiting each other with total mass $m=m_1+m_2$, mass ratio $\nu=m_2 m_1^{-1} \in (0,1)$ and symmetric mass ratio $\eta=\nu(1+\nu)^{-2} \in (0,0.25)$. According to the proposed model, the jet velocity results from a sum between the intrinsic jet launching velocity vector and the orbital velocity vector of the jet emitter BH \textit{at the ejection time} \citep{Roos1993,Kun2014,Kun2015}. According to the physical picture, while the general kinematics of the component can be explained by a helical jet, the oscillatory motion around this helical pattern reveals an additional kinematic influence manifesting itself as the oscillatory motion of the jet component, which we explain through an orbiting BBH.

\begin{figure*}
    \includegraphics[width=0.8\textwidth]{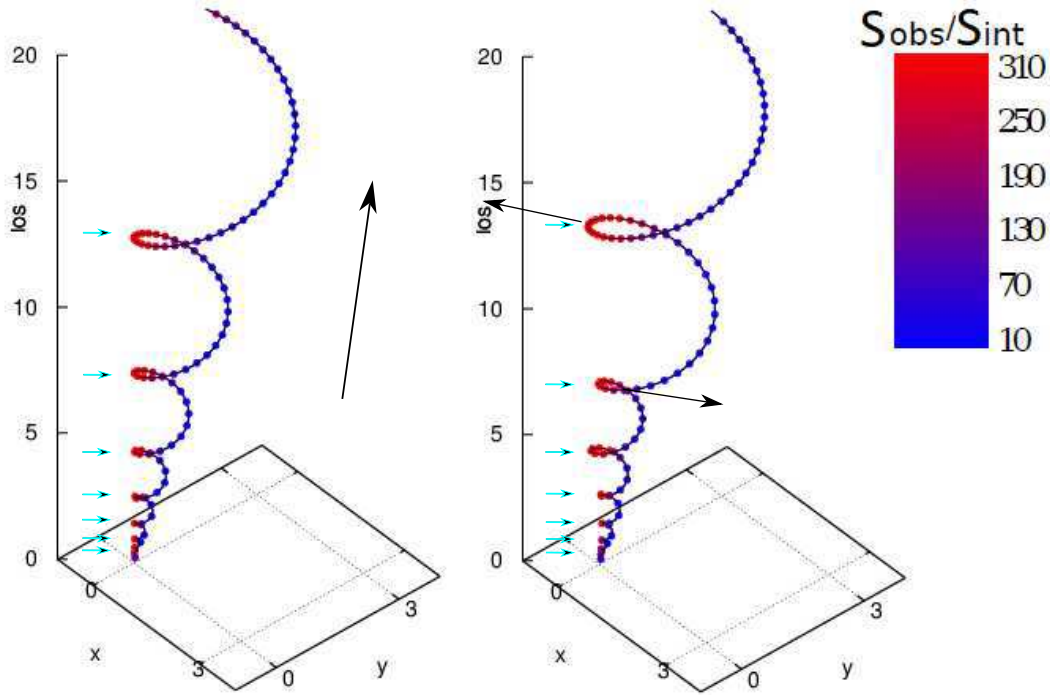}
    \caption{3D visualisation of the helical jet in the left panel, and its response to a passing wave in the right panel. The colouring of the particles corresponds to the experienced Doppler boost. The intrinsic brightness and the speed of the particles are constant along the ridge line, therefore the inhomogeneities in their apparent flux are fully accounted to their varying inclination. The largest arrow on the left panel points the global direction of the jet flow, while the large arrows on the right panel show the directions where the lantern regions moved during their oscillatory motion. The lantern regions are marked by filled turquoise arrows.}
    \label{fig:jetcomp}
\end{figure*}

\begin{table}
\caption{Simulation parameters.}
\label{table:simulation}
\begin{tabular}{lll}
\hline
\hline
Set & Parameter & Value \\  
\hline
\multirow{3}{*}{Run} & component number & $N=301$ \\
 & ejection time difference & $dt=0.2$ yr\\
 & total elapsed time & $t=60.2$ yr\\
 \hline
\multirow{2}{*}{Orientation} & position angle of jet axis & $\theta_0=40\degr$\\
 & inclination of jet axis & $\iota_0=5\degr$\\
 \hline
\multirow{6}{*}{Helical jet} & half-angle & $\psi_0=5\degr$\\
 & jet velocity & $\beta=0.9944 c$\\
 & initial angular velocity & $\omega_0=100$ mas yr$^{-1}$\\
 & initial core separation & $r_0=0.01$ mas\\
 & spectral index & $\alpha=-0.054$\\
 & geometry index & $n=2$\\
 \hline
\multirow{4}{*}{BBH}& total mass & $m=10^{10} M_{\odot}$\\
 & Newtonian orbital period & $T_N=10$ yr\\
 & mass ratio & $\nu=1/2$\\
 & spin angle & $\kappa=0\degr$\\
\hline
\end{tabular}
\end{table}

We show the simulation of a helical jet described by Eqs. (\ref{eq:rt}--\ref{eq:zt}), and its response to the orbiting nature of its base in $3$ dimension ($3$D) in Fig. \ref{fig:jetcomp}. The summary of the simulation parameters is given in Table \ref{table:simulation}. These parameters reflect the superluminal jet of a flat spectrum source, in order to visualize the jet-behaviour expected in such a physical picture, most importantly the flaring behaviour of the small-inclined jet curves (marked by arrows in Fig. \ref{fig:jetcomp}). The colouring of the jet corresponds to its apparent brightness, calculated as $S_\mathrm{app}/S_\mathrm{int}=\delta^{n+\alpha}$, where $S_\mathrm{int}$ is the intrinsic brightness chosen to be unity, $\delta$ is the Doppler factor, $\alpha$ is the spectral index of the jet, and $n=2\mathrm{, }3$ for continuous or discrete jet, respectively.

If the orbital velocity of the jet emitter BH is of the order of the intrinsic jet velocity, the jet launching direction will be imprinted with a secondary pattern due to the periodical changing direction of the BH velocity vector, and a wave-like pattern appears passing through the jet. The pitch of this structure is constant along the symmetry axis of the jet. According to the proposed model, the final jet ridge line is a vector summation of the intrinsically helical jet and this wave-like pattern. After the jet-ejection, the binary motion has no effect on the jet propagation. Another particular feature of the model is that the width of the jet changes in accordance with the phase of the passing wave. 

We call the areas where the jet shows maximum brightness boosting lantern regions. In the case of a non-orbiting jet base, presented in the left panel of Fig. \ref{fig:jetcomp}, the position and flux density of these lantern regions do not change. For an orbiting jet base on the other hand, presented in the right panel of Fig. \ref{fig:jetcomp}, the position and flux density of the lantern regions are altered by the passing wave-like perturbation inducing the flaring behaviour of the lantern regions. According to this scenario, as plasma reaches a lantern region, it apparently flares up, than after passing through the region it fades away. This cyclical alternation of the plasma explains why the components do not exhibit outward motion, rather maintain their globally stationary position as manifestation of the lantern regions. 



If we assume the wave passing through the jet is the consequence of the orbiting motion of the jet base, then the intrinsic orbital period emerges as $T=T_\mathrm{obs} (1+z)^{-1}\approx 6.8~\mathrm{yr} (1+0.679)^{-1}=4.05~\mathrm{yr}$. 
Consequently the orbital separation is
\begin{equation}
r=0.0057\times \left(\frac{m}{10^8 m_\mathrm{Sun}}\right)^{1/3} \mathrm{(pc)}
\end{equation}
or using the angular scale for S5 1803+784 ($6.917$ pc mas$^{-1}$)
\begin{equation}
r=0.0008\times \left(\frac{m}{10^8 m_\mathrm{Sun}}\right)^{1/3} \mathrm{(mas)}.
\end{equation}
The spin-orbit precession emerges as
\begin{equation}
T_\mathrm{SO}=2418\times \left(\frac{m}{10^8 m_\mathrm{Sun}}\right)^{-2/3} \frac{(1+\nu)^2}{\nu} \mathrm{(yr)},
\end{equation}
and the inspiral time as:
\begin{equation}
T_\mathrm{merger}=4.96\times10^6\times \left(\frac{m}{10^8 m_\mathrm{Sun}}\right)^{-5/3} \frac{(1+\nu)^2}{\nu} \mathrm{(yr)}.
\end{equation}

Using the independent black hole mass estimation $m_\mathrm{BH}\approx3.98\times 10^8~\mathrm{m_{\odot}}$ of \citet{Sbarrato2012}, for the binary separation to Newtonian order we find $r\approx0.009~\mathrm{pc}$, with the post-Newtonian parameter $\varepsilon=Gm r^{-1}c^{-2}\approx0.002$. If we set the low (high)-limit on the typical mass ratio of the SMBH mergers, i.e. $\nu=1/30$ ($1/3$) \citep{Gergely2009}, we find a spin-precession period $T_\mathrm{SO}\approx 3.1\times10^4$ yr ($5.1\times10^3$ yr), and an inspiral time $T_\mathrm{insp}\approx15.9\times 10^6$ yr ($2.6\times10^6$) yr. 

\section{Discussion}
\label{discussion}

In Fig. \ref{fig:coresep_all} we presented the integrated flux density, core separation, position angle and FWHM size of components B3, C0, C1, Ca, C2, C4, C8 and C12. We also mark the epochs when minima were found in their $x$-positions. 
The scenario of a passing wave shaking up the jet suggests the position and the brightness of the jet components should be correlated. We performed a Pearson-type correlation analysis between their core separation ($r$), integrated flux density ($S$), position angle ($\theta$), and FWHM size ($d$), in order to explore possible correlations between them. We calculated the correlation-coefficients for those components that appeared at least in $10$ epochs (C$0$, C$1$, C$a$, C$2$, C$4$, C$8$). The bootstrap method has been applied to calculate the expected value and the standard deviation of $R$ regarding the given component. The results of the correlation analysis are summarized in Table \ref{table_correlation}, where we marked by boldface the strongest cases. The strongest ($R\approx-0.64$) and most significant ($4\sigma<\alpha$) correlation emerged between the integrated flux density and the core separation of component C$2$. Mildly strong ($0.4\lesssim\mid R \mid$) and significant ($2\sigma\leqslant \alpha$) correlation is found between the flux density and the core separation of components Ca, the flux density and the position angle of C4 (see Table \ref{table_correlation}). The absence of stronger correlations might imply the flux variation of the lantern regions have possibly a composite origin due to inhomogeneities of the physical properties of the plasma, and of its Doppler boosting. In the former case the flux variability is not necessarily a consequence of structural evolution of the jet.

\begin{table*}
\caption{$C_\mathrm{ID}$: component \textit{ID}, $R$($r$,$S$): correlation-coefficient between $r$ and $S$, $\alpha$($r$,$S$): significance of $R$($r$,$S$), $R$($\theta$,$S$): correlation-coefficient between $\theta$ and $S$, $\alpha$($\theta$,$S$): significance of $R$($\theta$,$S$), $R$($S$,$d$): correlation-coefficient between $S$ and $d$, $\alpha$($S$,$d$): significance of $R$($S$,$d$). $2\sigma\leqslant$ significance levels are indicated by boldface.}
\label{table_correlation}
\begin{tabular}{c|r|r|r|r|r|r}

$C_\mathrm{ID}$ & $R$($r$,$S$)  & $\alpha$($r$,$S$) &  $R$($\theta$,$S$) &  $\alpha$($\theta$,$S$)& $R$($S$,$d$) &  $\alpha$($S$,$d$)\\
\hline
C$0$ & $-0.274 \pm 0.036$ & $0.9$     & $ 0.017 \pm 0.069$ & $<0.75$ & $ 0.315 \pm 0.049$ & $\mathbf{0.95}$ \\
C$1$  & $ 0.019 \pm 0.047$ & $<0.75$   & $ 0.280 \pm 0.046$ & $0.9$   & $ 0.284 \pm 0.048$  & $0.9$ \\
C$a$  & $ 0.401 \pm 0.700$ & $\mathbf{0.98}$ & $-0.268 \pm 0.035$ & $0.9$ & $ 0.447 \pm 0.056$ & $\mathbf{0.99}$ \\
C$2$  & $-0.636 \pm 0.044$ & $\mathbf{>0.9995}$ & $-0.056 \pm 0.037$ & $<0.75$ & $-0.304 \pm 0.092$ &  $0.9$ \\
C$4$  & $-0.033 \pm 0.098$ & $<0.75$   & $-0.428 \pm 0.068$ & $\mathbf{0.98}$  &  $0.438 \pm 0.080$ &  $\mathbf{0.99}$ \\
\hline
\end{tabular}
\end{table*}

We offered a scenario that explains qualitatively the oscillatory motion of the VLBI jet of the BL Lac object S5 1803+784, the orbital motion of the jet base forcing to oscillate its components. While this scenario can explain orientation effects leading to variable apparent flux density, its possible intrinsic origin needs to be further investigated. The possibility of a binary black hole (BBH) lying at the jet base of S5 1803+784 has been also discussed in \citet{Roland2008}. They solved first an astrometric problem by using the VLBI coordinates to provide the kinematical-parameters, such that the inclination $5.3{\degr}^{+1.7}_{-1.8}$, and the Lorentz factor $\gamma=3.7{\degr}^{+0.3}_{-0.2}$ emerged. Then they fitted their model and provided the BBH parameters giving the best fit to the kinematical data of S5 1803+784, first by assuming equal BH masses, then with variable mass ratio. In the non-equal mass case, with a dominant mass of $7.14\times 10^8 M_\odot$, the binary period was $T_b=1299.2 \mathrm{yr}$. This is much longer than the intrinsic period $T$ determined from our analysis.

In earlier decades, periodic jet structures were often used to be attributed to the orbital motion of SMBHs. However, hydrodynamic and later magneto-hydrodynamic simulations proved instabilities in the plasma also could lead to a helical structure in the jet \cite[see references for both physical pictures in][]{Kun2014}. The purpose of the creation of such a jet kinematical model then the presented one targets this ambiguity, aiming to soften the tension between the different models. According to the presented physical picture, while a helical jet can explain the general kinematics of the jet-components, this helical pattern reveals an additional kinematic influence due to the presence of an orbiting SMBH when its orbital speed is at least several per-cent of the jet speed.

The major difference between our model and the one presented by \cite{Roland2008} lies in the concepts of the underlying jet kinematics. While in the model of Roland et al. the component-motion was explained by the mixed effect of the precession of the accretion disk and the motion of the two SMBHs, in our model the jet can be naturally helical. \textit{If} there is no binary SMBH harboured at the centre if the AGN, the jet ridge-line can be still curved, as presented in the left side panel of Fig. \ref{fig:av_jetshape}. In that sense our model does not contradict with the expectations based on the HD/MHD simulations, namely that instabilities in the plasma can cause the jet to bend. Roland's model lacks this feature, and this is the reason why we updated the binary-concept of S5 1803+784 with the present qualitative model.

Furthermore, in \citet{Roland2008}, the identification of the components was different than in both previous and subsequent papers. For example in its Figs. 4-6., the C0 and C1 components are fitted together, like these two would be actually one VLBI component. Furthermore, we used a model in which the intrinsic period of the oscillation was assumed to be the orbital period. In contrast to \citet{Roland2008}, who incorporated the jet orientation by solving an astrometrical problem, we deduced its inclination from the Doppler boosted light of the core, and the superluminal speed of the only outwardly moving component $B3$.

We proposed that the jet-components are regions exhibiting flaring behaviour due to the oscillation of the ridge line about Doppler boosted regions. One could ask how B3 fits this physical picture. In B2010 the authors found the spectral behaviour of S5 1803+784 changed significantly after $\sim1996$, where the spectra changed from being flat, and gradually become steeper (at $4.6$, $8$, $15$ GHz). The changes of the spectral evolution occur in the beginning of a bright, prolonged flare C that started in $\sim1992$ and ended in $\sim2005$ with a peak in $1997$. They suggested that this can be explained by the ejection of the jet component B3 at launching time $1999.8$. Possibly born in an extremely violent ejection, we speculate that B3 is the only jet-component which is an actual self-consisting bunch of plasma, the brightness of which makes it visible on its full trajectory.

SMBH binaries, coalescing presumably on time scale of several decades, are potential sources of the low frequency gravitational waves (GWs) to be detected by the future LISA space mission. Identification of sources is important for breaking degeneracies in GW's parameter estimation, as well as to help calibrate the expected abundance of GW sources. Both the study of \citet{Roland2008} and the present one suggest \textit{if} S5 1803+784 indeed harbours a BBH, its inspiral time is more than one million years. This time-scale is too long to consider this AGN as possible source of strong GW bursts to be detected for example by the future LISA space mission. 

\section{Summary}
\label{section:summary}

In this work we have studied the jet of the blazar S5~1803+784 based on the VLBI data by the MOJAVE survey at the observational frequency 15 GHz. For that purpose we have model-fitted the calibrated $uv$-data obtained between 2005.85 and 2012.91. We also made use the results of B2010, which were derived based on the observation-period of $1994.67$--$2005.68$, also taken from the MOJAVE survey. Below we summarize our main findings and results.
\begin{enumerate}[label=\roman*,align=CenterWithParen]
\item We have found that the new data supports the oscillatory motion of the components proposed in B2010 based upon $11$ years of data ($1994.67$--$2005.68$).
\item The only superluminal component, B3 slowed down and faded out after $2011$.
\item We have characterized the time-scale of the oscillation, which emerged to be about six years.
\item The excess variance of the positional variability suggests the jet components being farther from the VLBI core have larger amplitude in their position variations.
\item The fractional variability amplitude shows slight changes in $3$-year bins of the component's position.
\item The average inclination of the VLBI jet emerged as $\iota_0\approx1.7\degr$ at 15 GHz observing frequency, and its average apparent width as $dP_\mathrm{av}\approx14\degr$.
\item The jet ridge line can be fitted with the helical jet model of \citet{Steffen1995}, which is based on the conservation laws for kinetic energy, momentum and jet opening angle.
\item Assuming the oscillatory motion of the jet of S5~1803+784 is a consequence of the presence of an orbiting SMBH at the jet-base, we characterized the Newtonian binary parameters.
\item The correlation-coefficients between $r$ and $S$ suggest, that the flux variation of the lantern regions have possibly a composite origin due to inhomogeneities of the physical properties of the plasma, and of its Doppler boosting.
\end{enumerate}

The main findings of the paper are (i) the stationarity of most of the components and the evolution of the jet ridge line; (ii) the identification of special lantern regions along the jet ridge line. Their average position marks the intrinsic jet shape, while their oscillatory motion is a manifestation of a perturbation passing through the jet. The alternation of the plasma flowing through the lantern regions explain why the jet-components do not exhibit outward motion, rather maintain their globally stationary position. All of these findings emphasize the role of orientation effects when studying jets with small inclination.

\section*{Acknowledgements}
E. K. thanks to G. Sz\H{u}cs for the valuable discussions. E. K. and L. \'{A}. G. acknowledge the support of the Hungarian National Research, Development and Innovation Office (NKFI) in the form of the grant 123996.
K. \'{E}. G. was supported by the J{\'a}nos Bolyai Research Scholarship of the Hungarian Academy of Sciences, and by
the Hungarian National Research Development and Innovation Office (OTKA NN110333). O.M.K acknowledges financial support by the Shota Rustaveli NSF under contract FR/217554/16.
This research has made use of data from the MOJAVE database that is maintained by the MOJAVE team (Lister et al., 2009, AJ, 137, 3718).

\end{document}